\documentclass[runningheads]{llncs}

\usepackage[T1]{fontenc}
\usepackage[utf8]{inputenc}
\usepackage{amsmath}
\usepackage{booktabs}
\usepackage{graphicx}
\usepackage{microtype}
\usepackage{multirow}
\usepackage{url}
\usepackage{float}
\usepackage{hyperref}
\usepackage{placeins}
\usepackage[shortcuts,acronym]{glossaries}

\newacronym{vni}{VNI}{Visual Networking Index}
\newacronym{ism}{ISM}{Industrial, Scientific, and Medical}
\newacronym{5g}{5G}{Fifth Generation}
\newacronym{6g}{6G}{Sixth Generation}
\newacronym{b5g}{B5G}{Beyond 5G}
\newacronym{m2m}{M2M}{Machine-to-Machine}
\newacronym{pid}{PID}{Proportional–Integral–Derivative}
\newacronym{pi}{PI}{Proportional–Integral}
\newacronym{cpu}{CPU}{Central Processing Unit}
\newacronym{ram}{RAM}{Random Access Memory}
\newacronym{sla}{SLA}{Service Level Agreement}
\newacronym{slo}{SLO}{Service Level Objective}
\newacronym{fifo}{FIFO}{First In, First Out}
\newacronym{Sim-Diasca}{Sim-Diasca}{Simulation of Discrete Systems of All Scales}
\newacronym{thd}{THD}{Threshold}
\newacronym{dc}{DC}{Data Center}
\newacronym{e2e}{E2E}{End-to-End}
\newacronym{6g-twin}{6G-TWIN}{6G-TWIN}
\newacronym{ce}{CE}{Computing Element}
\newacronym{ne}{NE}{Network Element}
\newacronym{iot}{IoT}{Internet of Things}
\newacronym{rps}{RPS}{Requests per Second}
\newacronym{http}{HTTP}{Hypertext Transfer Protocol}
\newacronym{lan}{LAN}{Local Area Network}
\newacronym{keda}{KEDA}{Kubernetes Event-driven Autoscaling}
\newacronym{hpa}{HPA}{Horizontal Pod Autoscaler}
\newacronym{rr}{RR}{Round Robin}
\newacronym{oran}{O-RAN}{Open Radio Access Network}
\newacronym{itut}{ITU-T}{International Telecommunication Union Telecommunication Standardization Sector}
\newacronym{post}{POST}{HTTP POST method}

\newacronym{ran}{RAN}{Radio Access Network}
\newacronym{upf}{UPF}{User Plane Function}
\newacronym{vran}{vRAN}{Virtualized Radio Access Network}
\newacronym{cn}{CN}{Core Network}
\newacronym{tn}{TN}{Transport Network}
\newacronym{ns}{NS}{Network Services}
\newacronym{vn}{VN}{Virtual Network}
\newacronym{vnf}{VNF}{Virtual Network Function}
\newacronym{pnf}{PNF}{Physical Network Function}
\newacronym{cnf}{CNF}{Cloud-Native Network Function}
\newacronym{vm}{VM}{Virtual Machine}
\newacronym{sdn}{SDN}{Software Defined Networking}
\newacronym{nfv}{NFV}{Network Function Virtualization}
\newacronym{mano}{MANO}{Management and Orchestration}
\newacronym{mec}{MEC}{Multi-access Edge Computing}
\newacronym{sfc}{SFC}{Service Function Chaining}
\newacronym{ue}{UE}{User Equipment}
\newacronym{wifi}{Wi-Fi}{IEEE 802.11}
\newacronym{wlan}{WLAN}{Wireless Local Area Network}
\newacronym{ap}{AP}{Access Point}
\newacronym{sta}{STA}{Station}
\newacronym{csma}{CSMA/CA}{Carrier-Sense Multiple Access with Collision Avoidance}
\newacronym{mcs}{MCS}{Modulation and Coding Scheme}
\newacronym{csa}{CSA}{Channel Switch Announcement}
\newacronym{cb}{CB}{Channel Bonding}
\newacronym{dcb}{DCB}{Dynamic Channel Bonding}
\newacronym{scb}{SCB}{Static Channel Bonding}
\newacronym{rssi}{RSSI}{Received Signal Strength Indicator}
\newacronym{sinr}{SINR}{Signal-to-Interference plus Noise Ratio}
\newacronym{obss}{OBSS}{Overlapping Basic Service Set}
\newacronym{bss}{BSS}{Basic Service Set}
\newacronym{sr}{SR}{Spatial Reuse}
\newacronym{dcf}{DCF}{Distributed Coordination Function}
\newacronym{ofdma}{OFDMA}{Orthogonal Frequency Division Multiple Access}
\newacronym{cca}{CCA}{Channel Clear Assessment}
\newacronym{ed}{ED}{Energy Detection}

\newacronym{lte}{LTE}{Long-Term Evolution}
\newacronym{lte-u}{LTE-U}{LTE Unlicensed}
\newacronym{lte-laa}{LTE-LAA}{LTE Licensed Assisted Access}
\newacronym[longplural={Transmission Opportunities}]{txop}{TXOP}{Transmission Opportunity}

\newacronym{sdr}{SDR}{Software Defined Radio}
\newacronym{usrp}{USRP}{Universal Software Radio Peripheral}
\newacronym{cr}{CR}{Cognitive Radio}
\newacronym{imt}{IMT}{International Mobile Telecommunications}
\newacronym{cu}{CU}{Centralized Unit}
\newacronym{du}{DU}{Distributed Unit}
\newacronym{nh}{NH}{Neutral Host}
\newacronym{nhn}{NHN}{Neutral Host Network}
\newacronym{mno}{MNO}{Mobile Network Operator}

\newacronym{ai}{AI}{Artificial Intelligence}
\newacronym{ml}{ML}{Machine Learning}
\newacronym{dl}{DL}{Deep Learning}
\newacronym{lr}{LR}{Linear Regression}
\newacronym{dtr}{DTR}{Decision Tree Regression}
\newacronym{dnn}{DNN}{Deep Neural Network}
\newacronym{xgboost}{XGBoost}{eXtreme Gradient Boosting}

\newacronym{rl}{RL}{Reinforcement Learning}
\newacronym{drl}{DRL}{Deep Reinforcement Learning}
\newacronym{dqn}{DQN}{Deep Q-Network}
\newacronym{pomdp}{POMDP}{Partially Observable Markov Decision Process}
\newacronym{mdp}{MDP}{Markov Decision Process}
\newacronym{ddpg}{DDPG}{Deep Deterministic Policy Gradients}
\newacronym{ppo}{PPO}{Proximal Policy Optimization}
\newacronym{a2c}{A2C}{Advantage Actor Critic}
\newacronym{trpo}{TRPO}{Trust Region Policy Optimization}
\newacronym{aco}{ACO}{Ant Colony Optimization}
\newacronym{pso}{PSO}{Particle Swarm Optimization}
\newacronym{mlops}{MLOps}{Machine Learning Operations}

\newacronym{sl}{SL}{Supervised Learning}
\newacronym{cnn}{CNN}{Convolutional Neural Network}
\newacronym{fnn}{FNN}{Feedforward Neural Network}
\newacronym{nn}{NN}{Neural Network}
\newacronym{mlp}{MLP}{Multilayer Perceptron}
\newacronym{ae}{AE}{Autoencoder}
\newacronym{relu}{ReLU}{Rectified Linear Unit}
\newacronym{gb}{GB}{Gradient Boosting}
\newacronym{svm}{SVM}{Support Vector Machine}
\newacronym{rf}{RF}{Random Forest}
\newacronym{dt}{DT}{Decision Tree}
\newacronym{lstm}{LSTM}{Long Short-Term Memory}

\newacronym{gnn}{GNN}{Graph Neural Network}
\newacronym{gcn}{GCN}{Graph Convolutional Network}
\newacronym{gcl}{GCL}{Graph Convolutional Layer}
\newacronym{gnb}{gNB}{gNodeB}
\newacronym{sb3}{SB3}{Stable-Baselines3}

\newacronym{ndt}{NDT}{Network Digital Twin}
\newacronym{tndt}{TNDT}{Telemetry-driven Network Digital Twin}
\newacronym{tdl}{TDL}{Telemetry Data Layer}
\newacronym{hdl}{HDL}{Harmonization Data Layer}
\newacronym{dt-c}{DT-C}{Digital Twin Connector}
\newacronym{amqp}{AMQP}{Advanced Message Queuing Protocol}
\newacronym{dds}{DDS}{Data Distribution Service}
\newacronym{mqtt}{MQTT}{Message Queuing Telemetry Transport}
\newacronym{coap}{CoAP}{Constrained Application Protocol}
\newacronym{nats}{NATS}{Neural Autonomic Transport System}
\newacronym{udr}{UDR}{Unified Data Repository}

\newacronym{qos}{QoS}{Quality of Service}
\newacronym{qoe}{QoE}{Quality of Experience}
\newacronym{rmse}{RMSE}{Root Mean-Squared Error}
\newacronym{kpi}{KPI}{Key Performance Indicator}
\newacronym{wfq}{WFQ}{Weighted Fair Queuing}
\newacronym{drr}{DRR}{Deficit Round Robin}
\newacronym{red}{RED}{Random Early Detection}

\newacronym{itu}{ITU}{International Telecommunication Union}
\newacronym{etsi}{ETSI}{European Telecommunications Standards Institute}

\newacronym{zsm}{ZSM}{Zero-Touch Network and Service Management}
\newacronym{cli}{CLI}{Command Line Interface}
\newacronym{snmp}{SNMP}{Simple Network Management Protocol}
\newacronym{nf}{NF}{Network Function}
\newacronym{k8s}{K8s}{Kubernetes}
\newacronym{mae}{MAE}{Mean Absolute Error}
\newacronym{bmp}{BMP}{BGP Monitoring Protocol}
\newacronym{int}{INT}{In-band Network Telemetry}
\newacronym{ds}{DS}{Data Space}
\newacronym{sdm}{SDM}{Smart Data Models}
\newacronym{ioc}{IOC}{Information Object Classes}
\newacronym{ddsp}{DDSP}{Distributed Data Stream Processing}
\newacronym{amf}{AMF}{Access and Mobility Management Function}

\usepackage{eso-pic}

\usepackage{tikz} 

\def\BibTeX{{\rm B\kern-.05em{\sc i\kern-.025em b}\kern-.08em
    T\kern-.1667em\lower.7ex\hbox{E}\kern-.125emX}}

\newcommand\copyrighttext{%
  \footnotesize This is the author's version of an article that has been accepted at the 9th International Conference on Machine Learning for Networking (MLN 2026), to be held on December 1--3, 2026 .}
\newcommand\copyrightnotice{%
\begin{tikzpicture}[remember picture,overlay]
\node[anchor=north,yshift=0pt] at (current page.north) {\fbox{\parbox{\dimexpr0.9\textwidth-\fboxsep-\fboxrule\relax}{\copyrighttext}}};
\end{tikzpicture}%
}

\begin{document}

\title{Stress2Twin: Profile-Driven Network Digital Twins for Guarded Edge Computing Resource Management}
\titlerunning{Stress2Twin: Profile-Driven NDTs for Edge Resource Management}

\author{Julian Jimenez Agudelo\inst{1} \and
Paola Soto\inst{1} \and
Ayat Zaki-Hindi\inst{2} \and
S\'ebastien Faye\inst{2} \and
Nina Slamnik-Krije\v{s}torac\inst{1} \and
Johann Marquez-Barja\inst{1} \and
Miguel Camelo Botero\inst{1}\thanks{Corresponding author.}}

\authorrunning{Julian Jimenez Agudelo et al.}

\institute{University of Antwerp -- IMEC, IDLab, Belgium\\
\email{miguel.camelobotero@uantwerpen.be} \and
Luxembourg Institute of Science and Technology (LIST), Luxembourg}

\maketitle
\copyrightnotice
\begin{abstract}
Heterogeneous edge systems require resource managers to determine how many application replicas to deploy, where to place them, and how to distribute incoming requests while satisfying service-level performance constraints. Acquiring the performance knowledge required for these decisions through exhaustive multi-node characterization is measurement-intensive and becomes increasingly difficult as the number of workers and configurations grows. Stress2Twin is a profile-driven \gls{ndt} framework that profiles edge nodes independently and composes their measured behavior into a system-level resource-management model. Machine-level oracles characterize latency, throughput, loss, queueing, and power as functions of assigned request rate and replica count. A guarded optimizer derives operating regions and tests neighboring request-rate states before retaining a replica count, reserving capacity before measured nonlinear degradation. Independent profiling of our three-worker Kubernetes testbed requires 271 measured states, corresponding to approximately 4.5 hours, compared with an estimated 123 hours for exhaustive characterization of the considered cluster configuration space. In a single-run ramp evaluation, the selected policy records no latency or loss-threshold violations within the derived 1,650-\gls{rps} supported range and reduces estimated power by 27.84\% relative to \gls{rps}100 \gls{rr} placement. An additional single-run evaluation under a non-monotonic workload records no latency violation and 19.67\% lower estimated energy than a stabilized \gls{rps}50 \gls{rr} policy. These results show that independently acquired machine profiles can support capacity-aware resource consolidation within a measured operating range. Power savings use an idealized active-worker accounting model.
\end{abstract}


\glsresetall

\section{Introduction}

\Gls{mec} places computation close to connected devices. It supports latency-sensitive \gls{iot} analytics and \gls{ai}-enabled edge inference. Unlike homogeneous data centers, compact edge deployments may combine machines with different processing capacities, energy costs, and replica limits. Resource management in such environments therefore requires jointly deciding how many application replicas to deploy, where to place them, and how incoming application requests should be distributed across heterogeneous edge compute nodes. Making these decisions requires knowledge of how candidate configurations behave under different traffic loads, but acquiring this knowledge directly at the cluster level quickly becomes expensive. Recent work on zero-touch orchestration similarly highlights the need to coordinate service placement and scaling across heterogeneous edge and cloud infrastructures while considering performance and energy-related metrics~\cite{cuervobello2026zsm}.

\Glspl{ndt} offer a natural mechanism for representing such performance knowledge and evaluating candidate configurations before applying them to the physical infrastructure. \Gls{itut} Y.3090 defines an \gls{ndt} as a virtual representation that uses data, models, and interfaces to analyze, emulate, and control a physical infrastructure~\cite{itutY3090}. Prior work has applied digital twins to energy-aware edge migration~\cite{bozkaya2025migration}, emulated-to-physical \gls{oran} validation~\cite{yaqoob2025orbit}, \gls{qos}-aware vehicular data distribution~\cite{rosa2025qos}, and trustworthy counterfactual what-if analysis for edge resources~\cite{Agudeloeucnc2026}. However, these studies do not address how to derive joint scaling, service-placement, and request-routing policies from independently measured node-level behavior without exhaustive system-level characterization.

Stress2Twin is a profile-driven \gls{ndt} decision framework for resource management in heterogeneous edge systems. Rather than characterizing complete multi-node configurations, it profiles each edge compute node independently as a function of assigned application request rate and active application replicas, composes the resulting machine-level models, and derives joint scaling, placement, and routing decisions. Its objective is to identify resource-efficient operating configurations that satisfy service-specific latency and loss constraints while avoiding operation close to measured saturation. The model covers the computing and service-routing domain of the \gls{mec} infrastructure, not the complete communication network; radio-access conditions, time-varying transport delays, and network-topology dynamics remain outside the present scope.

Related resource managers obtain performance knowledge differently. Autopilot combines historical observations and heuristics for resource configuration~\cite{rzadca2020autopilot}, Quasar classifies workloads for \gls{qos}-aware allocation~\cite{delimitrou2014quasar}, and Ernest derives performance models from a limited number of experiments~\cite{venkataraman2016ernest}. These approaches motivate data-assisted resource management, but do not jointly derive replica counts, placement, and routing from independently measured profiles of heterogeneous edge workers. A central challenge is therefore how to acquire sufficient performance knowledge for these joint decisions without exhaustively profiling the complete cluster configuration space. Table~\ref{tab:related} distinguishes Stress2Twin from representative performance-aware resource managers and \gls{ndt} studies. Existing methods provide valuable historical, workload-classification, learned-performance, or what-if models, whereas Stress2Twin specifically combines independently measured heterogeneous-worker profiles with joint replica, placement, and routing decisions and an empirical transition guard.

\begin{table}[t]
\centering
\caption{Positioning against representative related approaches.}
\label{tab:related}
\begin{tabular}{@{}lcccc@{}}
\toprule
Approach & Knowledge & HW heterog. & Joint$^{a}$ & Guard \\
\midrule
Autopilot~\cite{rzadca2020autopilot} & History/heuristics & -- & -- & -- \\
Quasar~\cite{delimitrou2014quasar} & Classification & Yes & -- & -- \\
Ernest~\cite{venkataraman2016ernest} & Sparse experiments & -- & -- & -- \\
Trustworthy NDT~\cite{Agudeloeucnc2026} & Telemetry/models & Yes & -- & -- \\
Stress2Twin & Per-worker profiles & Yes & Yes & Yes \\
\bottomrule
\end{tabular}
\vspace{1pt}
\begin{minipage}{0.96\textwidth}\scriptsize
$^{a}$Joint replica count, worker placement, and request routing. A dash means that the cited work does not target that complete decision set.
\end{minipage}
\end{table}

Exhaustive cluster profiling becomes combinatorial as the number and capacity of workers increases, consistent with the data-acquisition and validation challenges reported for digital twins~\cite{fuller2020digital}. For $m$ workers with maximum replica counts $r_1,\ldots,r_m$, each traffic level already admits $\prod_i(r_i+1)-1$ nonempty replica configurations before routing is considered. For example, three workers supporting 8, 4, and 4 replicas therefore yield 224 configurations per offered-load level. Across 33 offered-load levels, measured in \gls{rps}, evaluating each configuration for 60 seconds would require approximately 123 hours. Repeating such exhaustive measurements whenever a worker, workload, or deployment characteristic changes further limits their practical use for constructing and maintaining an \gls{ndt}.

A more measurement-efficient alternative is to profile each machine independently over its supported combinations of assigned load and replica count and subsequently compose these measurements at the cluster level. Each retained profile state represents one measured worker-load-replica combination. For each replica count, the assigned load starts at 50~\gls{rps} and increases in 50-\gls{rps} steps every 60 seconds until observed request loss reaches or exceeds 50\%. Retaining only states with loss below 50\% yields 271 machine-level profile states across NUC0, NUC1, and NUC2.

For the 50-\gls{rps} profiling grid used in the ramp evaluation, each state is measured for approximately 60 seconds, giving 16{,}260 seconds, or about 4.5 hours, across the 271 retained states. This is substantially smaller than the approximately 123 hours required for exhaustive cluster-level profiling, because the proposed procedure scales with the sum of admissible single-machine states rather than with the product of cluster-wide configurations. This reduction introduces a second challenge: independent profiles characterize workers under isolated, steady-state operation and cannot be interpreted directly as safe cluster-level operating points.

Stress2Twin addresses this gap by composing replaceable machine models into a cluster-level \gls{ndt}. A guarded optimizer derives operating regions from the profiles, checks neighboring loads, and constructs coordinated replica, active-worker, and routing decisions under latency and loss constraints. This capacity-aware consolidation avoids the overload risk of load-only sequential scaling and the persistent worker activation of round-robin placement.

The main contributions of this work are threefold. First, Stress2Twin introduces a profile-driven \gls{ndt} that composes independently measured machine-level performance profiles into a cluster-level decision model, reducing the need for exhaustive characterization of the complete configuration space. Second, it proposes a guarded optimization mechanism that derives per-worker operating regions from latency, throughput, loss, and queue measurements and uses neighboring load states to construct coordinated replica-count, worker-activation, and routing decisions while avoiding operation close to nonlinear degradation. Third, it validates the resulting policies on a heterogeneous Kubernetes edge deployment, quantifying both simulation-to-reality accuracy and the trade-offs in latency violations, replica use, active workers, and estimated power against fixed \gls{rps} and latency-aware baselines, and supplements the controlled ramp with an evaluation under a non-monotonic workload.

The rest of the paper is structured as follows. Section~\ref{sec:approach} formulates the resource-management problem and presents the proposed approach. Section~\ref{sec:system} describes the experimental platform and measurement architecture. Sections~\ref{sec:functional} and~\ref{sec:optimization} define the functional model and profile-derived optimizer, respectively. Section~\ref{sec:evaluation} evaluates the resulting policies, while Sections~\ref{sec:discussion} and~\ref{sec:conclusion} discuss the findings and conclude the paper.

\section{Problem Formulation and Proposed Approach}
\label{sec:approach}

We consider a heterogeneous edge system composed of a set of compute nodes $\mathcal{N}$ that may differ in processing capacity, energy consumption, and the number of application instances they can host. We refer to each independently runnable instance of the application as a \emph{replica}. The offered load $\Lambda$ denotes the incoming application request rate. For a given $\Lambda$, the resource manager must determine three coupled decisions: how many replicas to deploy, on which edge compute nodes they should run, and what fraction of the incoming requests should be routed to each active node. The objective is to satisfy service-specific latency and loss constraints while avoiding unnecessary replica and node activation and, consequently, limiting resource and power consumption.

A central difficulty is obtaining the performance knowledge required to evaluate candidate configurations. Direct cluster-level characterization requires testing combinations of replica counts, worker placements, traffic allocations, and offered loads. The number of candidate configurations therefore increases rapidly with the number and capacity of workers, and repeating this characterization after changes in hardware, software, or workload further increases the measurement cost.

Stress2Twin addresses this problem by moving performance characterization from the cluster level to individual workers. Each worker is profiled independently over supported combinations of assigned load and replica count. These measurements are represented through a machine-level oracle that estimates the worker's latency, loss, throughput, queue behavior, and power for a candidate operating point. The cluster-level \gls{ndt} subsequently composes the outputs of the individual machine oracles to estimate the behavior of candidate multi-worker configurations.

Independent steady-state profiles alone, however, are insufficient to guarantee safe operation. A measured state may satisfy the final service-level constraint while operating close to a saturation transition, where a small increase in assigned traffic can produce a disproportionate increase in latency or loss. Stress2Twin therefore derives guarded operating regions from the measured profiles and evaluates not only the current candidate state but also neighboring load states. The resulting optimizer constructs coordinated replica-count, active-worker, and routing decisions while retaining reserve capacity before measured nonlinear degradation.

The approach thus separates machine-specific behavior, which can be replaced or updated when the hardware or application changes, from cluster-level policy generation over the composed machine models. The following section instantiates this formulation in the heterogeneous Kubernetes testbed.

\section{System and Measurement Architecture}\label{sec:system}

\subsection{Edge \gls{ai} Deployment}

\begin{figure}[H]
    \centering
    \includegraphics[width=1.0\textwidth]{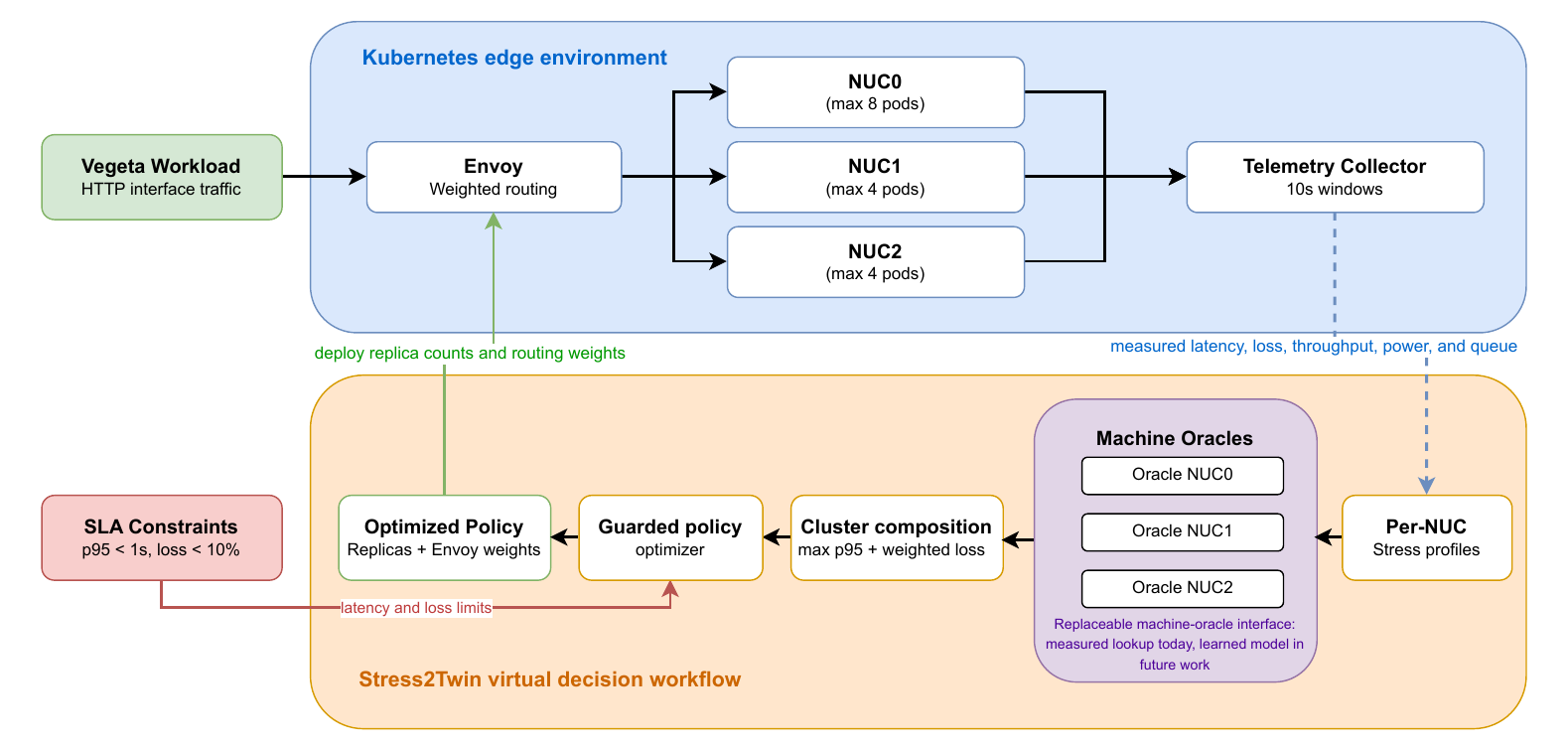}
    \caption{Stress2Twin architecture. Per-worker measurements create replaceable machine oracles, which are composed to evaluate cluster configurations under latency and loss constraints. The guarded optimizer returns replica counts and Envoy routing weights to the Kubernetes deployment.}
    \label{fig:deployment}
\end{figure}

The system contains one Kubernetes control-plane node and three inference workers, denoted NUC0, NUC1, and NUC2 (Fig.~\ref{fig:deployment}). The control-plane node coordinates the experiment and hosts the request endpoint, but runs no inference pods. NUC0 is the strongest measured worker and supports at most eight inference pods. NUC1 and NUC2 support four pods each. The resulting 8-4-4 limit follows the available hardware threads, with one runnable worker context per pod. Table~\ref{tab:platform} summarizes the compute platform.

\begin{table}[H]
\centering
\caption{Experimental compute platform.}
\label{tab:platform}
\begin{tabular}{@{}llll@{}}
\toprule
Node & Processor & Cores/threads & Memory \\
\midrule
Control & Intel Core i3-1115G4 & 2/4 & 16 GB \\
NUC0 & Intel Core i7-8650U & 4/8 & 8 GB \\
NUC1 & Intel Core i5-7260U & 2/4 & 32 GB \\
NUC2 & Intel Core i5-7260U & 2/4 & 8 GB \\
\bottomrule
\end{tabular}
\end{table}

The application running in the cluster is a containerized service for CIFAR-10 image classification~\cite{krizhevsky2009cifar}. Each replica is allocated one thread and 1 GB of memory. Envoy distributes requests using weights $W_1$, $W_2$, and $W_3$ for NUC0, NUC1, and NUC2. A policy specifies the replicas and routing weight on every worker. The optimizer activates workers according to a fixed capacity-ordered consolidation policy, filling NUC0 first and subsequently activating NUC1 and NUC2 as the offered load exceeds the guarded capacity of the active workers.

Figure~\ref{fig:deployment} shows the physical deployment and the Stress2Twin digital-twin workflow. Following the ITU-T Y.3090 framing of an \gls{ndt} as a virtual representation built from data, models, and interfaces, the functional model in Section~\ref{sec:functional} provides Stress2Twin's model component. The physical system comprises the Kubernetes cluster, its edge workers, and the traffic-serving and monitoring components.

Measurements collected for each worker across combinations of offered load and active replica count are aggregated into replaceable machine oracles. The functional model composes these oracles to estimate cluster-level performance for candidate configurations. The guarded optimizer then uses those estimates to select replica counts, active workers, and Envoy routing weights. In this way, the twin connects physical-system measurements to resource-management decisions and supports evaluating configurations without exhaustively testing each one on the cluster. The implementation uses measured lookup oracles populated from the profiling workflow.

\subsection{Telemetry and Phase Definition}

Vegeta runs on a separate client machine on the same \gls{lan}. Each emulated client randomly selects a CIFAR-10 image from the complete dataset of 60,000 images, serializes it, and sends a POST inference request over a keep-alive \gls{http} connection. The retained request trace reports a median request size of 2,492 bytes and median response size of 1,622 bytes. The \gls{k8s} deployment uses a custom \textit{cifar-server} container image derived from a Python base image, with Envoy HTTPRoute weights and a 15-second route timeout; no application retry policy is recorded. The kube-prometheus-stack scrapes the service and collector. Kepler supplies node-energy indicators, while Grafana supports inspection. A dedicated collector aggregates 10-second windows, persists them in InfluxDB v2, and publishes them through a Kafka topic for Prefect-managed \gls{ndt} workflows~\cite{Agudeloeucnc2026}. The evaluated deployment uses \gls{k8s} 1.35.7. Raw requests and phase summaries yield successful-request p95 latency, loss, throughput, queue indicators, replicas, and power. Loss is one minus the successful-request fraction; observed failures comprise \gls{http} 503 responses and transport timeouts.

\begin{figure}[!tbp]
    \centering
    \includegraphics[width=\textwidth]{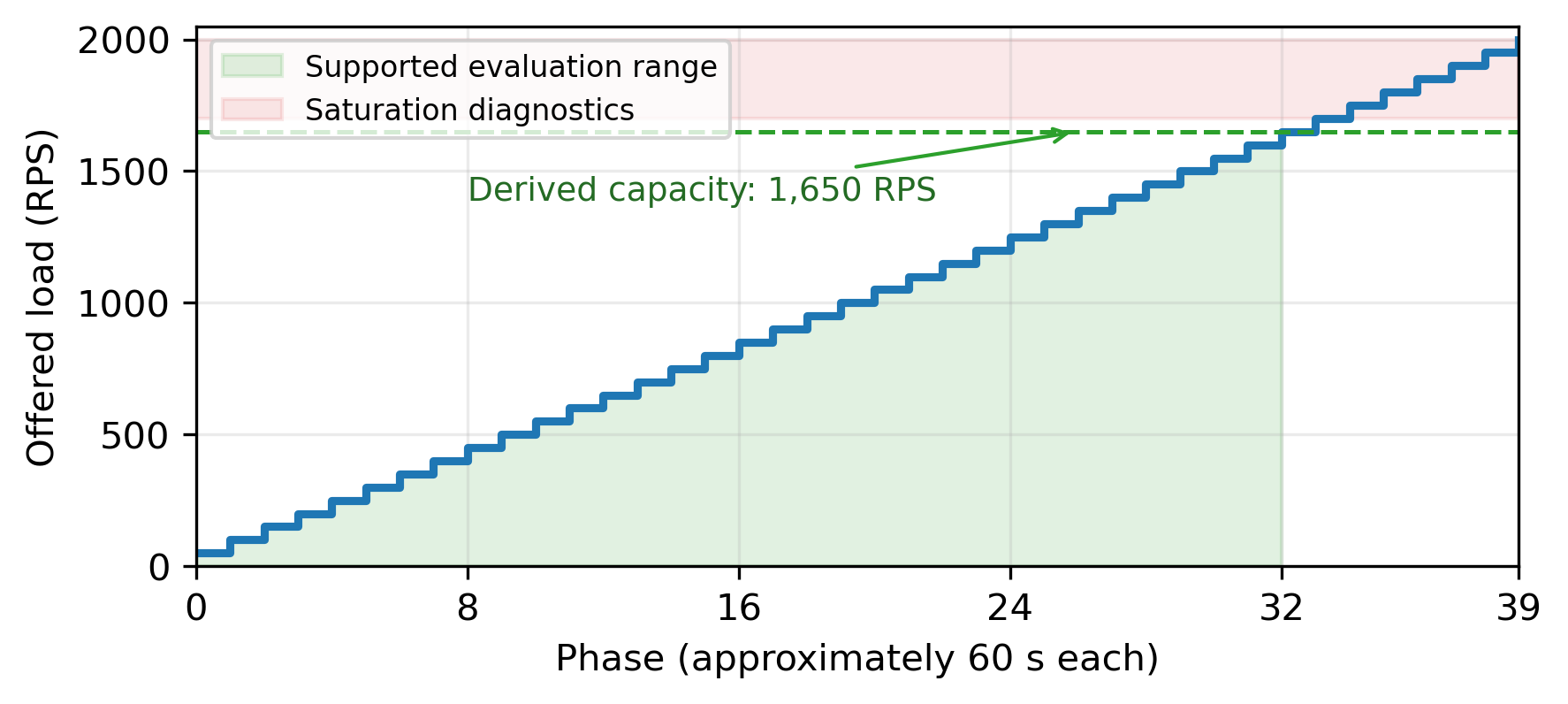}
    \caption{Forty-phase evaluation ramp and the profile-derived supported range.}
    \label{fig:workload}
\end{figure}

A \emph{phase} is a constant-load interval of approximately 60 seconds. The evaluation consists of 40 phases indexed by $k=0,\ldots,39$, with offered load $\Lambda_k=50(k+1)$ \gls{rps}. Thus, the workload increases in 50-\gls{rps} steps from 50 to 2,000 \gls{rps}. At each phase, the replica counts and routing weights associated with the corresponding offered load are applied. Figure~\ref{fig:workload} shows the resulting workload ramp. Phases up to 1,650 \gls{rps} constitute the supported comparison range, while loads from 1,700 to 2,000 \gls{rps} are retained as saturation diagnostics and excluded from performance denominators.

\subsection{Power Accounting}

The physical workers were not physically switched off in any trace. We therefore estimate the intended deployment power by counting a worker whenever the evaluated policy keeps it active:
\begin{equation}
\begin{aligned}
\widehat{P}_i &= 0, && \text{if $i$ is inactive},\\
\widehat{P}_i &= 11\,\mathrm{W}+
\max(P_{\mathrm{K},i}-P_{\mathrm{K0},i},0),
&& \text{otherwise}.
\end{aligned}
\label{eq:power}
\end{equation}
The 11-W term is an independently measured idle wall-power value, with a standard deviation of 0.3 W. Here, $P_{\mathrm{K},i}$ is the Kepler estimate for worker $i$ at the evaluated state, and $P_{\mathrm{K0},i}$ is its node-specific no-load Kepler reference. The second term retains the load-dependent Kepler component above that reference. Cluster power is the sum across active workers. Since phases have equal duration, differences in average estimated power are proportional to differences in estimated energy over the evaluated trace.

\section{Stress2Twin Functional Model}\label{sec:functional}

\subsection{Machine-Level Oracle}

\begin{figure}[!tbp]
    \centering
    \includegraphics[width=1.0\textwidth]{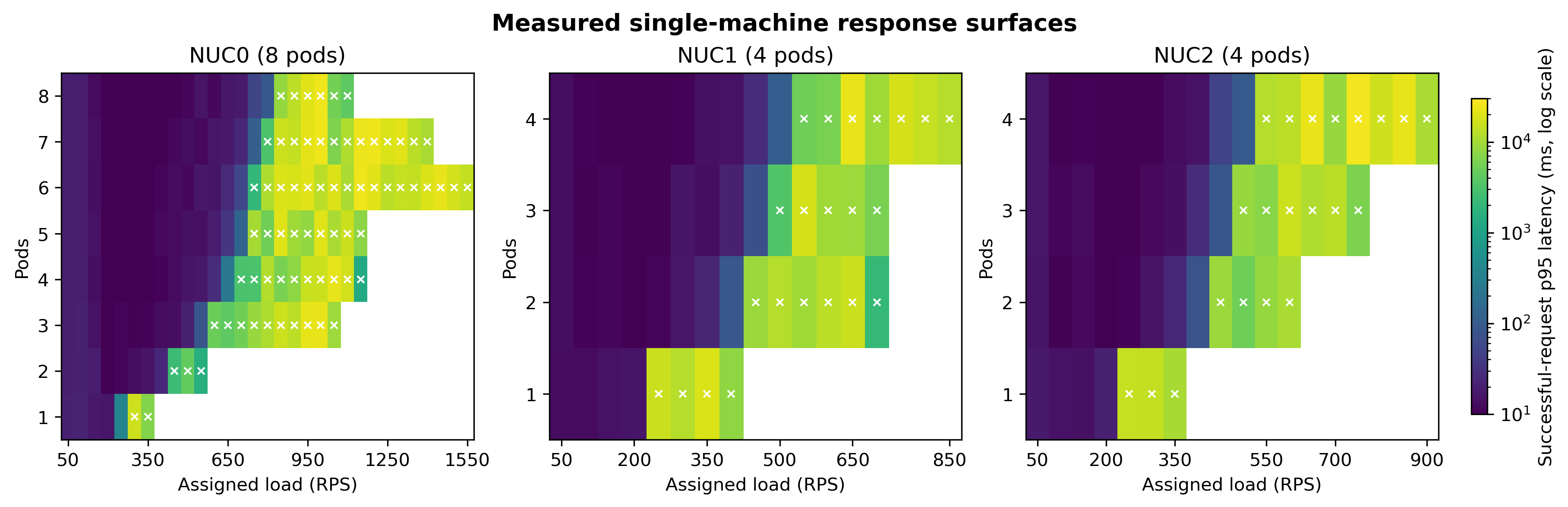}
    \caption{Measured p95-latency profiles for NUC0, NUC1, and NUC2. White crosses identify states that violate the one-second latency \gls{sla}.}
    \label{fig:profiles}
\end{figure}

Each worker is stress-tested independently over measured combinations of assigned load $\lambda_i$ and active replicas $n_i$ (Fig.~\ref{fig:profiles}). Request-level measurements are aggregated by worker, load, and replica count. For each state, p95 is computed directly over its successful request latencies rather than by averaging 10-second-window percentiles; the collector windows provide the queue and energy summaries. The resulting oracle implements
\begin{equation}
f_i(\lambda_i,n_i)\rightarrow
\{p95_i,\ell_i,\mu_i,P_i,q_i^{95},d_i^{95}\},
\label{eq:oracle}
\end{equation}
where $p95_i$ is successful-request latency, $\ell_i$ is request loss, $\mu_i$ is successful throughput, $P_i$ is the load-dependent Kepler power component used in Eq.~(\ref{eq:power}), and $q_i^{95}$ and $d_i^{95}$ are the p95 queueing time and queue depth, respectively. The complete estimated active-worker power is $\widehat{P}_i$. The interface is independent of the implementation: the lookup oracle can later be replaced by a learned model without changing the optimizer. For policy construction, the optimizer derives its curves from measured profiling-grid states. During simulation, it maps an unavailable load to the nearest measured load at the same replica count, considering another replica count only when the requested count has no measured state.

\subsection{Cluster Composition}

For a candidate whose assigned loads sum to cluster load $\Lambda$, the cluster predictions are composed conservatively:
\begin{equation}
p95_{\mathrm{cluster}}=\max_i p95_i,
\qquad
\ell_{\mathrm{cluster}}=
\frac{\sum_i\lambda_i\ell_i}{\Lambda}.
\label{eq:composition}
\end{equation}
The maximum latency prevents a lightly weighted but overloaded worker from being hidden by averaging. Loss is request-weighted because each worker receives a different share of offered traffic. A feasible candidate must satisfy $p95_{\mathrm{cluster}}<1{,}000$ ms and $\ell_{\mathrm{cluster}}<10\%$.

Latency and loss thresholds are service-specific inputs. Applying Stress2Twin to another service requires profiling it on the target hardware and deriving the corresponding operating-region boundaries and guarded capacities.

\section{Profile-Derived Policy Optimization}\label{sec:optimization}

\subsection{Steady-State Optimizer}

The steady-state optimizer enumerates measured worker states whose assigned loads sum to the requested cluster load. It rejects configurations that decrease a worker's load or replica count as the policy is constructed over ascending load levels. Candidates are ranked lexicographically by constraint violation, number of active workers, total replicas, and power. Consequently, configurations satisfying both service constraints are preferred, and the remaining objectives select among those feasible configurations. This procedure produces efficient operating points from the individual profiles. It defines the unguarded baseline; the guarded policy subsequently derives per-worker replica curves and assembles them using the capacity-ordered activation rule in Section~\ref{sec:lookahead}. The resulting policy curve is monotonic over ascending load levels. Under variable traffic, the current load selects its associated configuration, with the downscale hold of Section~\ref{sec:variable} governing replica reductions.

\subsection{Empirical Operating Regions}

The guarded optimizer distinguishes three operating regions. The \emph{stable region} contains low-latency states with adequate throughput and limited queueing. The \emph{shoulder} is the transition region in which latency and queue indicators rise but service has not collapsed. The \emph{collapse region} contains states after the largest distributional separation, where latency or loss increases sharply and throughput ceases to track offered load. These regions describe each worker's measured load-replica performance space. An operating point is defined by assigned request rate and active replica count, while measured latency, throughput, loss, and queue indicators determine its region.

Boundaries are thresholds that separate the three regions. They are obtained from the 271 eligible profile states, without using cluster-policy outcomes. First, the largest adjacent gap in $\log(1+p95)$ separates the shoulder from collapse. Second, within the pre-collapse states, the adjacent p95 gap around the 100-ms stability anchor separates stable states from the shoulder. The anchor selects the observed gap, whose midpoint defines the boundary. The two p95 thresholds use geometric midpoints because their gaps are identified on a logarithmic scale; the throughput threshold uses the arithmetic midpoint of its adjacent linear-scale gap. The calculations retain the measured endpoints:
\begin{equation}
\begin{aligned}
b_{\mathrm{stable}} &= \sqrt{(98.0+1)(111.0+1)}-1=104.3\ \mathrm{ms},\\
b_{\mathrm{collapse}} &= \sqrt{(372.38398+1)(1{,}346.20916+1)}-1\\
&=708.243\ \mathrm{ms},\\
r_{\min} &= (0.9549+0.9807)/2=0.9678.
\end{aligned}
\label{eq:boundaries}
\end{equation}
Queue guards are the upper limits observed in stable states, while outer queue limits are obtained from pre-collapse states. Loss remains the experiment constraint rather than a fitted boundary. Unlike Kneedle~\cite{satopaa2011kneedle}, which finds one knee on an ordered curve, this procedure separates regions across a discrete, multi-metric state set. Table~\ref{tab:boundaries} reports both the evidence and the resulting thresholds.

\begin{table}[H]
\centering
\caption{Profile-derived operating-region boundaries.}
\label{tab:boundaries}
\begin{tabular}{@{}lll@{}}
\toprule
Boundary & Observed separation & Threshold \\
\midrule
Stable/shoulder p95 & 98.0--111.0 ms & 104.3 ms \\
Shoulder/collapse p95 & 372.4--1,346.2 ms & 708.2 ms \\
Queue-time guard/outer & Upper stable/pre-collapse limits & 47.2/153.6 ms \\
Queue-depth guard/outer & Upper stable/pre-collapse limits & 1/4 requests \\
Throughput ratio & 0.9549--0.9807 & 0.9678 \\
Loss & Experiment constraint & $<10\%$ \\
\bottomrule
\end{tabular}
\end{table}

For each worker, guarded capacity is the largest assigned load for which some admissible replica count passes the current-state guard and neighboring-state outer-envelope tests. Applying Table~\ref{tab:boundaries} gives 750 \gls{rps} for NUC0 and 450 \gls{rps} for each of NUC1 and NUC2. Their sum defines the 1,650-\gls{rps} supported region. The fixed activation order follows the measured capacities and consolidation objective. NUC0 is filled first, followed by NUC1 and NUC2.

\subsection{Two-Step Neighborhood Test}\label{sec:lookahead}
For a given worker, the replica count $n$ determines how many parallel instances of the application are available to process its assigned load $\lambda$. Increasing $n$ provides additional processing capacity and can reduce queueing and latency as the offered load grows, at the cost of additional compute resources. Conversely, retaining too few replicas as $\lambda$ approaches the worker's capacity can move the worker toward the shoulder or collapse region. The neighborhood test therefore determines whether the current replica count can be safely retained as load increases, rather than scaling replicas only after an SLA violation is observed.

For measured load $\lambda$ and candidate replica count $n$, the selected optimizer examines the current state and the next two 50-\gls{rps} profile points. It retains $n$ only if the current point is in the stable guard region and both neighboring points remain inside the outer pre-collapse envelope:
\begin{equation}
\begin{split}
\operatorname{keep}(n,\lambda)\Leftrightarrow&\;\operatorname{guard}(n,\lambda)\land\\
&\operatorname{outer}(n,\lambda+50)\land
\operatorname{outer}(n,\lambda+100).
\end{split}
\label{eq:lookahead}
\end{equation}

If the current replica count does not satisfy the neighborhood test, the optimizer increases the number of replicas and selects the smallest larger count that provides a safe operating point for the current and neighboring loads. This avoids unnecessary overprovisioning while reserving sufficient processing capacity before the worker reaches the measured saturation region. The resulting per-worker replica curves are constrained to be monotonic. The evaluated guarded policy does not search arbitrary placements: at cluster level it activates workers according to the fixed capacity-ordered consolidation policy and computes integer routing weights in proportion to assigned load. The lookahead operates in load space: distances of 50 and 100 \gls{rps} correspond to one and two points on the profiling grid.

\section{Experimental Evaluation}\label{sec:evaluation}

\subsection{Baselines and Metrics}

The real-deployment evaluation includes two rule-based baseline families. Fixed-\gls{rps} policies allocate one replica per 50 or 100 \gls{rps}. \Gls{rr} keeps all three workers active throughout the experiment, places successive replicas cyclically, and weights traffic in proportion to active replicas. The worker count for these baselines therefore denotes workers retained active by the policy, not only workers hosting a replica in a particular phase. Sequential activation fills NUC0, then NUC1, then NUC2, but has no measured-capacity guard. Both fixed-\gls{rps} policies are capped at the 16-replica platform limit. The cap binds above 800 \gls{rps} for \gls{rps}50 and at 1,650 \gls{rps} for \gls{rps}100; without it, \gls{rps}50 would average 17.0 replicas rather than 12.36, and \gls{rps}100 would request 17 rather than 16 replicas in the final phase. The latency-aware \gls{keda} controller\footnote{\Gls{keda} Prometheus scaler: \url{https://keda.sh/docs/scalers/prometheus/}.} evaluates a 30-second Prometheus p95 every 10 seconds and scales when it exceeds 500 ms, providing a nominal 50\% margin before the one-second \gls{sla}. It uses one to 16 replicas, a 15-second service-monitor interval, and a 300-second cooldown, implemented as downscale stabilization. Its trace does not contain sufficient per-worker routing state to report average active workers. Native \gls{cpu}-driven Kubernetes \gls{hpa}\footnote{Kubernetes \gls{hpa}: \url{https://kubernetes.io/docs/tasks/run-application/horizontal-pod-autoscale/}.} was configured but did not trigger, and no aligned \gls{cpu} trace was retained, so it is excluded.

Stress2Twin is a workload-indexed policy. For fixed-\gls{rps} and Stress2Twin deployments, the offered phase load selects a precomputed configuration that is applied 15 seconds before each phase boundary, with a 120-second readiness timeout. \Gls{keda} detects degradation through a trailing window. The fixed-\gls{rps} policies therefore provide the direct configuration-level comparison, while \gls{keda} represents a reactive knowledge-based controller.

We report average active replicas, average active workers, estimated power, phases with p95 latency above one second, aggregate request loss, and phases with loss above 10\%. Aggregate loss is the fraction of all offered requests in the comparison interval that do not complete successfully. All comparisons use the same 33 phases from 50 to 1,650 \gls{rps}.
The evaluation is organized around four questions. First, does neighborhood protection improve the deployed policy relative to steady-state selection? Second, how closely do oracle predictions match deployment outcomes inside and outside the supported range? Third, how does the selected policy compare with rule-based scalers in service compliance and resource use? Finally, does the same profile-driven approach remain useful under a non-monotonic workload?
To support reproducibility, the datasets, analysis code, and scripts used to generate the reported results will be released in Zenodo.\footnote{\url{https://zenodo.org/records/23804238}.}

\subsection{Optimizer Ablation}

Table~\ref{tab:ablation} compares three levels of transition protection while retaining the same profile source, composition rule, and activation order. The steady-state optimizer uses the fewest replicas but records 21 p95 violations. In the corresponding single runs, a one-step neighborhood records one violation, and the second step records none for 0.39 additional average replicas relative to one-step lookahead. Average power remains essentially unchanged because reserve replicas are normally added to workers that are already serving.

\begin{table}[H]
\centering
\caption{Ramp-deployment optimizer ablation to 1,650 \gls{rps}.}
\label{tab:ablation}

\begin{tabular}{@{}lrrrr@{}}
\toprule
Optimizer & Replicas & Power (W) & p95 $>1$ s & Loss $>10\%$ \\
\midrule
Steady-state profile & 7.30 & 48.51 & 21 & 0 \\
One-step guarded & 8.03 & 52.41 & 1 & 0 \\
Two-step guarded & 8.42 & 52.31 & 0 & 0 \\
\bottomrule
\end{tabular}
\end{table}

Static profiles identify useful operating zones, but the final \gls{sla} is too permissive as a selection boundary. For the observed ramp, reserving capacity before the knee is associated with progressively fewer violations as the neighborhood increases from zero to two steps. The contrast between the policies shows that final-\gls{sla} feasibility and guarded feasibility lead to materially different real-deployment behavior, although repeated runs are required to quantify variability.

\subsection{Predicted Policy Outcomes and Real Deployment}\label{sec:simreal}

\begin{figure}[H]
    \centering
    \includegraphics[width=1.0\textwidth]{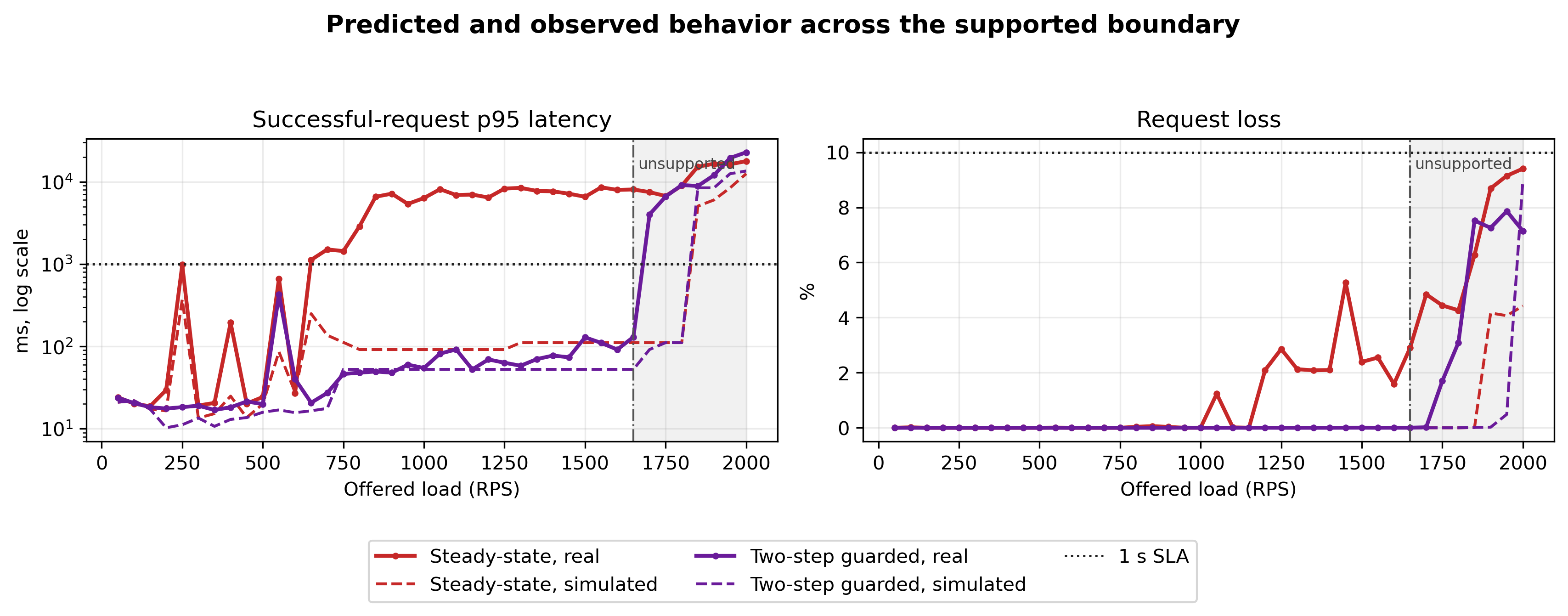}
    \caption{Predicted and observed latency and loss. The shaded region exceeds the derived 1,650-\gls{rps} supported range.}
    \label{fig:simreal}
\end{figure}

Figure~\ref{fig:simreal} compares oracle predictions with the outcome obtained after deploying the same policies. Table~\ref{tab:simreal} separates the supported region from the seven saturated phases. Within the supported region, the guarded policy has a p95 \gls{rmse} of 75.8 ms and a prediction bias of $-27.2$ ms. Although small relative to the one-second \gls{sla}, 75.8 ms is material relative to the tens-of-milliseconds latency of stable states. The log-p95 \gls{rmse} is 0.305 decades, equivalent to a typical multiplicative error factor of 2.02. Its loss \gls{mae} is 0.0005 percentage points and its power \gls{rmse} is 2.12 W. These errors are conditional on the configurations selected by each policy and do not compare oracle accuracy at identical operating states.

\begin{table}[H]
\centering
\caption{Prediction error by operating range. Bias is mean predicted minus observed latency; false-safe means predicted below both \gls{sla} limits but violated in deployment.}
\label{tab:simreal}
\footnotesize
\begin{tabular}{@{}llrrrr@{}}
\toprule
Policy & Range & p95 \gls{rmse} & Bias & Log \gls{rmse} & False-safe \\
 & & (ms) & (ms) & (decades) & phases \\
\midrule
Steady-state & 50--1,650 & 5,241.3 & $-3{,}943.9$ & 1.411 & 21/33 \\
Two-step & 50--1,650 & 75.8 & $-27.2$ & 0.305 & 0/33 \\
Steady-state & 1,700--2,000 & 8,273.3 & $-8,084.6$ & 1.235 & 3/7 \\
Two-step & 1,700--2,000 & 6,377.9 & $-5,653.5$ & 1.172 & 3/7 \\
\bottomrule
\end{tabular}

\end{table}

A false-safe phase is one that the oracle predicts below both \gls{sla} limits but the deployment violates at least one. It therefore measures erroneous feasible decisions made from the oracle output. The guarded policy has no false-safe phase in the supported region. Its first violation occurs at 1,700 \gls{rps}, one 50-\gls{rps} grid step above the derived capacity, and all seven phases above 1,650 \gls{rps} violate an \gls{sla}. This observation is consistent with the summed worker capacities to within one grid step for this policy at full replica count.

\subsection{Rule-Based Baselines and Resource Efficiency}

Table~\ref{tab:policies} contrasts consolidation alone with policies that also use measured per-worker capacity.

\begin{table}[H]
\centering
\caption{Single-run ramp results to 1,650 \gls{rps}.}
\label{tab:policies}
\small
\begin{tabular}{@{}lrrrrrr@{}}
\toprule
Policy & Replicas & Active & Power (W) & p95 viol. & Loss (\%) & Loss viol. \\
\midrule
\gls{rps}50 \gls{rr} & 12.36 & 3.00 & 71.40 & 2 & 0.0704 & 0 \\
\gls{rps}50 sequential & 12.36 & 2.39 & 63.89 & 2 & 0.0377 & 0 \\
\gls{rps}100 \gls{rr} & 8.73 & 3.00 & 72.49 & 2 & 0.0446 & 0 \\
\gls{rps}100 sequential & 8.73 & 1.79 & 50.74 & 14 & 0.3210 & 0 \\
Latency-aware 500 ms$^{a}$ & 3.73 & -- & $\sim$65.76 & 12 & 6.4716 & 8 \\
Steady-state profile & 7.30 & 1.73 & 48.51 & 21 & 1.3741 & 0 \\
Two-step guarded & 8.42 & 1.82 & 52.31 & 0 & 0.0008 & 0 \\
\bottomrule
\end{tabular}
\vspace{1pt}
\begin{minipage}{0.96\textwidth}\scriptsize
$^{a}$Per-worker routing was unavailable, so the power estimate assumes all three workers were kept active.
\end{minipage}
\end{table}

\Gls{rps}100 sequential activates only 1.79 workers on average and reaches the lowest power among the fixed-\gls{rps} policies, 50.74 W, but records 14 p95 violations while filling NUC0 without respecting its 750-\gls{rps} guarded capacity. \Gls{rps}100 \gls{rr} records two violations but keeps all workers active and consumes 72.49 W. With similar replica use, the two-step policy records no phase above either \gls{sla} threshold and consumes 52.31 W, 27.84\% less than \gls{rps}100 \gls{rr}.

\Gls{rps}50 sequential provisions about 1.4 times as many replicas as \gls{rps}100 sequential but still records two violations. The two-step policy uses 31.86\% fewer replicas and 18.1\% less power than this baseline, with no threshold violation. At 16 replicas, replica-proportional 8:4:4 routing sends 50\% of traffic to NUC0, exceeding its 750-\gls{rps} guarded capacity above 1,500 \gls{rps}. The retained logs confirm that \gls{rps}50 sequential, \gls{rps}50 \gls{rr}, and \gls{rps}100 \gls{rr} first violate p95 at 1,600 \gls{rps}. Thus, additional replicas are not sufficient when routing overloads one worker.

The results are consistent with capacity-aware consolidation that coordinates replicas, placement, and routing under service-level constraints. The latency-aware trace records 12 latency and eight loss-threshold violations, consistent with degradation entering its trailing observation window before detection, scheduling, and readiness complete. More generally, replica count alone is not a sufficient proxy for serving capacity in a heterogeneous cluster: placement and routing must also reflect the measured capacity of the workers hosting those replicas.

Relative to conservative \gls{rps}50 \gls{rr}, the two-step policy reduces replicas by 31.86\%, active workers by 39.39\%, and estimated power by 26.74\%. This comparison quantifies the cost of overprovisioning. The comparison with \gls{rps}100 is more demanding: integer phase totals fall from 288 to 278 replicas, a 3.47\% reduction, while active-worker consolidation yields a 27.84\% reduction in estimated power. The estimated energy benefit is therefore associated mainly with placement and routing rather than replica reduction alone.

\subsection{Policy Evaluation under a Non-Monotonic Workload}\label{sec:variable}

\begin{figure}[H]
    \centering
    \includegraphics[width=\textwidth]{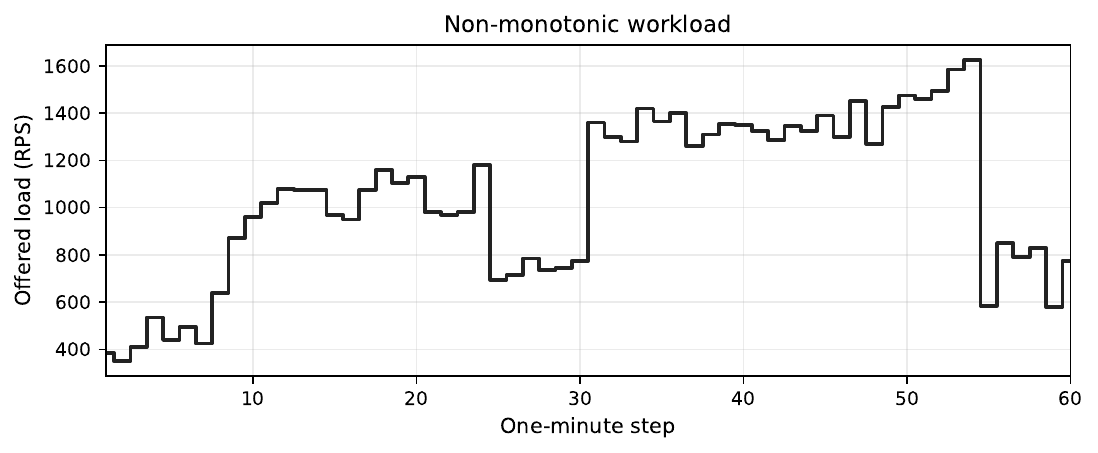}
    \caption{Non-monotonic offered-load sequence used for the 60-step deployment evaluation. Each step lasts one minute.}
    \label{fig:variable-workload}
\end{figure}

\begin{figure}[H]
    \centering
    \includegraphics[width=\textwidth]{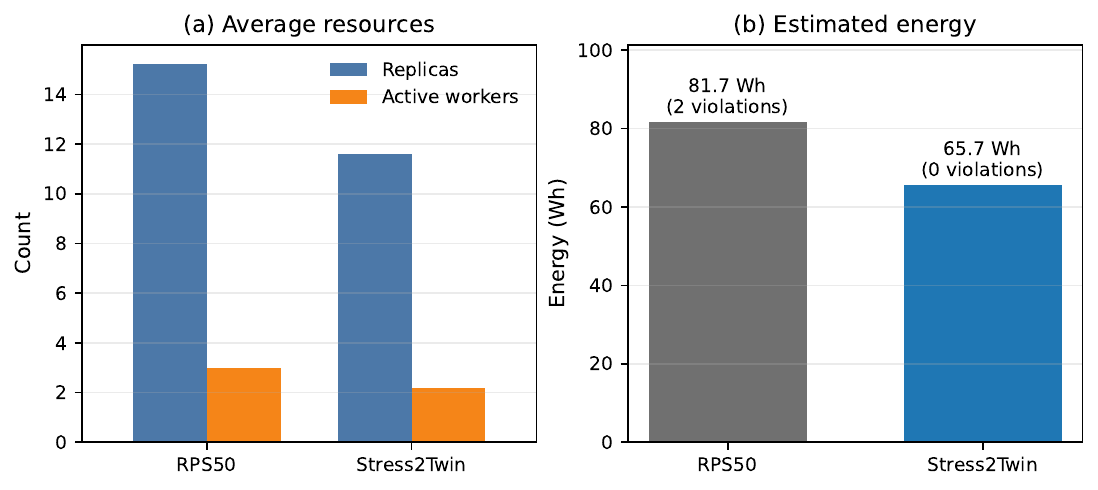}
    \caption{Average resource use and estimated energy under five-minute downscale stabilization. Energy labels also report the number of one-minute steps with p95 latency above one second.}
    \label{fig:variable-results}
\end{figure}

We additionally analyze a previously collected 60-step deployment trace. Each step lasts one minute, and offered load varies non-monotonically between 350 and 1,625~\gls{rps}. This separate Stress2Twin policy uses a denser 10-\gls{rps} profiling dataset and a configured 60-\gls{rps} neighborhood; it is not the 100-\gls{rps} two-step policy used for the ramp. The 271-state, 4.5-hour profiling cost reported above applies only to the 50-\gls{rps} ramp dataset.
Both Stress2Twin and the \gls{rps}50 \gls{rr} baseline retain the required replica count for five minutes before downscaling. This configuration evaluates the framework and its downscale stabilization under changing load. It supplements the controlled ramp rather than repeating the exact 50-\gls{rps}, two-step policy used above. Figure~\ref{fig:variable-workload} shows the offered-load sequence, while Fig.~\ref{fig:variable-results} and Table~\ref{tab:variable} report the resource and energy outcomes.

\begin{table}[H]
\centering
\caption{Single-run non-monotonic workload results.}
\label{tab:variable}
\begin{tabular}{@{}lrrrrrr@{}}
\toprule
Policy & Replicas & Active & Power (W) & Energy (Wh) & p95 viol. & Loss (\%) \\
\midrule
\gls{rps}50 \gls{rr}, 5-min hold & 15.23 & 3.00 & 81.42 & 81.74 & 2 & 0.0113 \\
Stress2Twin, 5-min hold & 11.62 & 2.20 & 65.53 & 65.66 & 0 & 0.0005 \\
\bottomrule
\end{tabular}
\end{table}

Across the complete trace, Stress2Twin reduces average replicas by 23.74\%, active workers by 26.67\%, average estimated power by 19.52\%, and estimated energy by 19.67\% relative to stabilized \gls{rps}50 \gls{rr}. It records no p95 violation, whereas the baseline records two; neither policy has a phase above 10\% loss.
In the Stress2Twin run without the hold, the policy uses 10.40 replicas on average and records three p95 violations; the stabilized run uses 11.62 replicas and records none. The comparison illustrates the resource-versus-transition-safety trade-off under decreasing as well as increasing load.

\section{Discussion and Practical Implications}\label{sec:discussion}

The results show that independently acquired machine profiles can support joint replica, placement, and routing decisions in the evaluated heterogeneous deployment. The contribution lies in how this performance knowledge is acquired and used: behavior is characterized per worker, composed conservatively at the cluster level, and filtered by operating boundaries derived from the profile states. Unlike resource managers that rely on historical cluster observations~\cite{rzadca2020autopilot,delimitrou2014quasar} or fit job-level performance models~\cite{venkataraman2016ernest}, Stress2Twin derives a policy without exhaustive cluster-level profiling, and its machine oracles can be replaced without modifying the optimizer.

\paragraph{Sources of the observed benefit.} The evaluated policies allow consolidation, transition protection, and capacity-aware routing to be partially separated. With identical replica schedules, \gls{rps}100 sequential reduces estimated power from 72.49 to 50.74~W relative to \gls{rps}100 \gls{rr}, but increases p95 violations from 2 to 14, consistent with consolidation moving load onto NUC0 beyond its guarded capacity. Between the two profile-derived policies, which share the same composition and routing logic, adding the neighborhood test coincides with a reduction from 21 violations to none.

The two-step policy keeps a similar number of workers active as \gls{rps}100 sequential (1.82 vs.\ 1.79) with fewer replicas (8.42 vs.\ 8.73) and records no violation, so its benefit is not explained by additional replicas. The \gls{rps}50 baselines confirm that more replicas do not restore safety when replica-proportional routing exceeds a worker's guarded capacity. The evaluation does not fully separate capacity-proportional routing weights from the replica reserve; a consolidation baseline that uses the guarded capacities without the neighborhood test would isolate these two components.

\paragraph{Energy accounting.} Reported power savings assume that workers not retained active by a policy consume no power, whereas the workers remained physically powered throughout the traces. An alternative idle-power scenario is obtained by charging every inactive worker its measured 11-W idle power. Because Eq.~(\ref{eq:power}) is linear in the active set, this adds $11\,(3-\bar{s})$~W to each policy, where $\bar{s}$ is its average number of active workers. Under these two accounting assumptions, the estimated power saving relative to \gls{rps}100 \gls{rr} ranges from 9.93\% to 27.84\%. This interval represents the sensitivity of the accounting model rather than a guaranteed bound on physically realizable savings. Actual wall-power reductions depend on the available power-state mechanisms, worker idle consumption, and boot and resume overheads.

\paragraph{Prediction error and the role of the guard.} The negative bias in Table~\ref{tab:simreal} shows that the steady-state oracle tends to underestimate deployed latency. The 21 false-safe phases of the steady-state policy are consistent with selecting isolated profile states that satisfy the final \gls{sla} but lie close to the shoulder, where small deviations in assigned load or queue state produce large latency changes. The lower error reported for the guarded policy is a selection effect: the guard excludes states where prediction discrepancies can change the feasibility decision, but it does not alter the oracle prediction at a fixed state. It is not a confidence interval and does not quantify predictive uncertainty. Since the remaining 75.8-ms p95 \gls{rmse} is material relative to stable latency, the twin should be used to rank configurations and delimit a safe region rather than to predict absolute latency. Propagating per-state measurement variability into explicit uncertainty margins is a natural extension.

\paragraph{Parameter selection.} The operating-region boundaries in Table~\ref{tab:boundaries} are derived from the 271 profile states without using cluster-policy outcomes, whereas the neighborhood depth was selected on the evaluation ramp. In the ablation runs, the violation count falls from 21 to one with the first step and from one to zero with the second, which adds 0.39 average replicas. This result motivates the selected depth but does not constitute out-of-sample tuning evidence. Deriving the depth from the profiles themselves, for example from the load distance between the guard and collapse regions of each worker, would remove this dependence and remains to be evaluated.

\paragraph{Workload dynamics and reactive control.} Stress2Twin is a workload-indexed policy, so it receives offered load, whereas the \gls{keda} controller infers degradation from a trailing latency window. The ramp comparison therefore quantifies the cost of reactive detection rather than ranking controllers with equal information; a load-driven \gls{keda} scaler would provide matched inputs. The non-monotonic trace extends the evidence to decreasing and irregularly changing load. The stabilized run records fewer violations than the unstabilized run, but single-run measurements cannot attribute that difference exclusively to the five-minute hold. Because the steady-state oracle omits backlog, startup, and readiness, this result does not establish robustness to arbitrary bursts or transition histories; those conditions require explicit dynamic state or conservative stabilization.

\paragraph{Scalability and repeatability.} Profiling cost grows with the sum of admissible per-worker states rather than with the product of cluster configurations, which keeps acquisition tractable as workers are added. Policy derivation scales differently: unrestricted enumeration of feasible cluster configurations grows combinatorially with the number of workers, whereas the evaluated guarded policy relies on per-worker profile lookups and a fixed capacity-ordered activation sequence, avoiding a search over arbitrary worker combinations. Its runtime and memory scaling beyond the evaluated configuration, and its behavior with many comparable workers, remain to be characterized. All results come from one run per policy; the reported counts and errors are therefore descriptive, and their variability across repeated runs remains to be established.

\paragraph{Validity of the operating envelope.} The 1,650-\gls{rps} boundary is a property of the measured hardware, application, and \gls{sla}, not a general capacity guarantee. The first violation of the guarded policy occurs one profiling-grid step above it, which supports its use as an operating envelope for this deployment. Profiles were acquired for a single service on a single \gls{lan}, so co-located services, resource contention, or transport variability would shift the measured boundaries. After material hardware, software, or workload changes, the affected workers must be re-profiled and their boundaries and guarded capacities re-derived; because profiling is per worker, only the changed workers need to be re-measured. 

\section{Conclusion and Future Work}\label{sec:conclusion}

Stress2Twin composes independent per-worker stress profiles into an \gls{ndt} decision model for heterogeneous \gls{mec} scaling. In the evaluated deployment, final-\gls{sla} feasibility alone was unsafe near nonlinear transitions, whereas a two-step, profile-derived neighborhood test records no phase above either the latency or loss threshold through the derived 1,650-\gls{rps} supported range. Against \gls{rps}100 \gls{rr}, it uses 27.84\% less estimated power, and against \gls{rps}100 sequential consolidation it records zero rather than 14 p95 violations at a power cost of 1.6 W, approximately 3\%. This outcome is consistent with per-worker capacity knowledge supporting consolidation without observed \gls{sla} violations. Under the same five-minute downscale hold, the non-monotonic evaluation records no p95 violation and 19.67\% lower estimated energy than \gls{rps}50 \gls{rr}.

Open challenges include validating repeatability and wall-power savings, deriving neighborhood depth without selecting it on the evaluation ramp and representing burst-driven transition dynamics, learning reusable per-worker models from sparse profiles, propagating prediction uncertainty, detecting model drift, and scaling policy search beyond exhaustive enumeration. Addressing them requires broader evaluation across traffic patterns, services, and hardware. Runtime residual monitoring could detect departures from the validated latency and loss envelope, enabling re-profiling or a conservative fallback.

\begin{credits}
\subsubsection{\ackname}
The Network Digital Twin modelling work was supported by the 6G-TWIN project, funded by the Smart Networks and Services Joint Undertaking under the European Union's Horizon Europe Research and Innovation Programme, Grant Agreement No. 101136314. The policy optimization algorithm work was supported by imec.icon RAPIDNESS, co-financed by imec and Flanders Innovation and Entrepreneurship (HBC.2024.0772).

\subsubsection{\discintname}
The authors have no competing interests to declare that are relevant to the content of this article.

\end{credits}

\bibliographystyle{splncs04}
\bibliography{sources}

\end{document}